\documentclass[a4paper, letter]{aa}
\usepackage{txfonts}
\usepackage{natbib}
\bibpunct{(}{)}{;}{a}{}{,}
\usepackage{graphics, color}
\begin{document}


\title{Optical polarimetry toward the Pipe nebula: Revealing the importance  
of the magnetic field\thanks{Based on observations collected at Observat\'orio
 do Pico dos Dias, operated by Laborat\'orio Nacional de Astrof\'\i sica 
(LNA/MCT, Brazil).}}

\author{F.~O. Alves\inst{1}
\and G.~A.~P. Franco\inst{2}
\and J.~M. Girart\inst{1}
}

\institute{
Institut de Ci\`encies de l'Espai (CSIC--IEEC), Campus UAB, Facultat de
Ci\`encies, C5 par 2$^{\rm a}$, 08193, Bellaterra, Catalunya, Spain \\
\email{[oliveira;girart]@ieec.uab.es}
\and
Departamento de F\'\i sica -- ICEx -- UFMG, Caixa Postal 702, 30.123-970 
Belo Horizonte, Brazil \\
\email{franco@fisica.ufmg.br}
}

\date{Received 29 April 2008/ Accepted 03 June 2008}

\titlerunning{The Pipe nebula}
\authorrunning{F.~O.~Alves et al.}

\abstract
{Magnetic fields are proposed to play an important role in the formation 
and support of self-gravitating clouds and the formation and evolution of 
protostars in such clouds.}
{We attempt to understand more precisely how the Pipe nebula is affected
by the magnetic field.} 
{We use $R$-band linear polarimetry collected for about 12\,000 stars in 
46 fields with lines of sight toward the Pipe nebula to investigate the 
properties of the polarization across this dark cloud complex.}
{Mean polarization vectors show that the magnetic field is locally
perpendicular to the large filamentary structure of the Pipe nebula (the 
`stem'), indicating that the global collapse may have been driven by ambipolar
diffusion. The polarization properties clearly change along the Pipe nebula. 
The northwestern end of the nebula (B59 region) is found to have a low 
degree of polarization and high dispersion in polarization position angle, 
while at the other extreme of the cloud (the `bowl') we found mean degrees of 
polarization as high as $\approx$15\% and a low dispersion in polarization 
position angle. The plane of the sky magnetic field strength was
estimated to vary from about 17\,$\mu$G in the B59 region to about 65\,$\mu$G
in the bowl.}
{We propose that three distinct regions exist, which may be related to 
different evolutionary stages of the cloud; this idea is supported by both 
the polarization properties across the Pipe and the estimated mass-to-flux 
ratio that varies between approximately super-critical toward the B59 region 
and sub-critical inside the bowl. The three regions that we identify are: the 
B59 region, which is currently forming stars; the stem, which appears to be 
at an earlier stage of star formation where material has been through a 
collapsing phase but not yet given birth to stars; and the bowl, which 
represents the earliest stage of the cloud in which the collapsing phase and
cloud fragmentation has already started.}


\maketitle

\section{Introduction}\label{int}

Understanding the role that magnetic fields play in the evolution of 
interstellar molecular clouds is one of the outstanding challenges of modern 
astrophysics. One problem related to star formation concerns the competition 
between magnetic and turbulent forces. The prevailing scenario of how stars
form is quasi-static evolution of a strongly magnetized core into a
protostar following influence between gravitational and magnetic 
forces. By ambipolar diffusion, i.e., the drift of neutral matter with 
respect to plasma and magnetic field, gravity finds a way to overcome 
magnetic pressure and eventually win the battle \citep[e.g., ][]{MS56, 
Na79, MP81, LS89}.  However, doubts about the validity of this theory were
expressed because of the apparent inconsistency between the expected and 
inferred lifetimes of molecular clouds. This inconsistency inspired
some researchers to propose a new theory in which star formation is 
driven by turbulent supersonic flows in the interstellar medium. Magnetic 
fields may be present in this theory, but they are too weak to be 
energetically important \citep[e.g. ][]{ES04,MK04}. It must be noted, 
however, that some results \citep{TM04,MTK06} demonstrate that the 
ambipolar--diffusion--controlled star formation theory is not in contradiction 
with molecular cloud lifetimes and star formation timescales.

\begin{figure*}[ht]
\sidecaption
\resizebox{12cm}{!}{\includegraphics{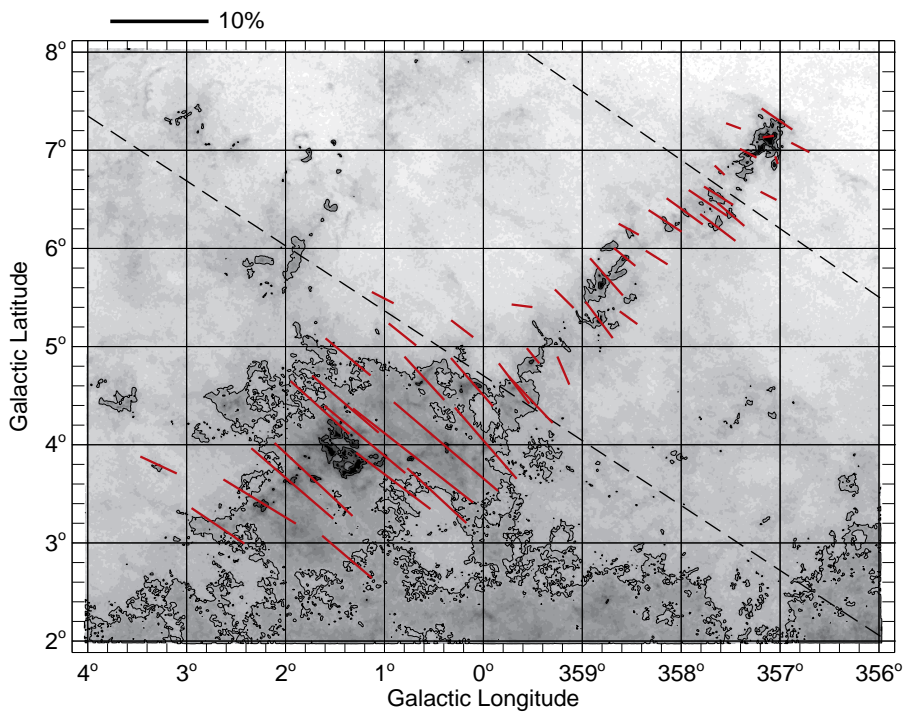}}
\caption[]{Mean polarization vectors, for each of the observed 46
fields, overplotted on the dust extinction map of the Pipe nebula obtained
by \citet{LAL06}. The lengths of these vectors are proportional to the
scale indicated in the top left-hand corner. Only stars showing
${\rm P}/\sigma_P \ge 10$ were used in the calculus of the mean
polarization and position angle. The dashed-lines indicate the celestial
meridians defined by $17^{\rm h} 14^{\rm m} 30\fs0$ and
$17^{\rm h} 27^{\rm m} 40\fs0$ (see text and Fig.\,\ref{fig2}).}
\label{fig1}
\end{figure*}

Previous optical polarimetric observations toward well-known forming molecular
clouds have enabled the large-scale magnetic field associated with these 
regions to be studied \citep[e.g.][]{GBM90}. In this work, we introduce the 
general results of a polarimetric survey conducted for the Pipe nebula, a 
nearby \citep[130--160\,pc,][]{LAL06, AF07} and massive 
($10^4\,{\rm M}_{\sun}$) dark cloud complex that appears to provide
a suitable laboratory for investigating magneto-turbulent phenomena. The 
Pipe nebula exhibits little evidence of star formation activity despite having
an appropriate mass. Until now, the only confirmed star-forming region in 
this nebula was B59 \citep{BHB07}, an irregularly-shaped dark cloud located at 
the northwestern end of the large filamentary structure that extends from 
($l,b$) $\approx$ ($0\degr, 4\degr$) to ($l,b$) $\approx$ 
($357\degr, 7\degr$). This apparently low efficiency in forming stars may 
be an indication of youth. \citet{ALL07} identified, in this cloud, 159 
cores of effective diameters between 0.1 and 0.4\,pc, and estimated 
masses ranging from 0.5 to 28 M$_{\sun}$, supposedly in a very early 
stage of development. A further investigation of these cores \citep{LMRAL} 
discovered that most of them appeared to be pressure confined and in 
equilibrium with the surrounding environment, and that the most massive
($\ga2$~M$_{\sun}$) cores were gravitationally bound. They suggested that the 
measured dispersion in internal core pressure of about a factor of 2--3 
could be caused by either local variations in the external pressure, or the 
presence of internal static magnetic fields with strengths of less than 16
\,$\mu$G, or a combination of both. The results derived from our optical 
polarimetric observations indicate that the magnetic field probably plays a 
far more important role in the Pipe nebula.

\section{Observations}\label{obs}

The polarimetric data were acquired using the 1.6\,m and the IAG 60\,cm 
telescopes of Observat\'orio do Pico dos Dias (LNA/MCT, Brazil) during
observing runs completed between 2005 to 2007. These data were acquired by
using a CCD camera specially adapted to allow polarimetric measurements; 
for a full description of the polarimeter see \citet{AM96}. $R$-band linear 
polarimetry, by means of deep CCD imaging, was obtained for 46 fields, each 
with a field of view of about $12'\times12'$, distributed over more than 
$7\degr$ (17~pc in projection) covering the main body of the Pipe nebula. 
The reference direction of the polarizer was determined by observing
polarized standard stars. For all observing seasons, the instrumental
position angles were perfectly correlated with standard values.
The survey contains polarimetric data of about 12\,000 stars, almost 6\,600 of
which have ${\rm P}/\sigma_P \ge 10$. The results presented in this Letter 
are based on the analysis of the latter group of stars. Details of 
observations, data reduction, and the analysis of the small-scale 
polarization properties within each observed area, will be described in a 
forthcoming paper \citep{FAG08}.

\begin{figure*}
\centering
\resizebox{16.5cm}{!}{\includegraphics{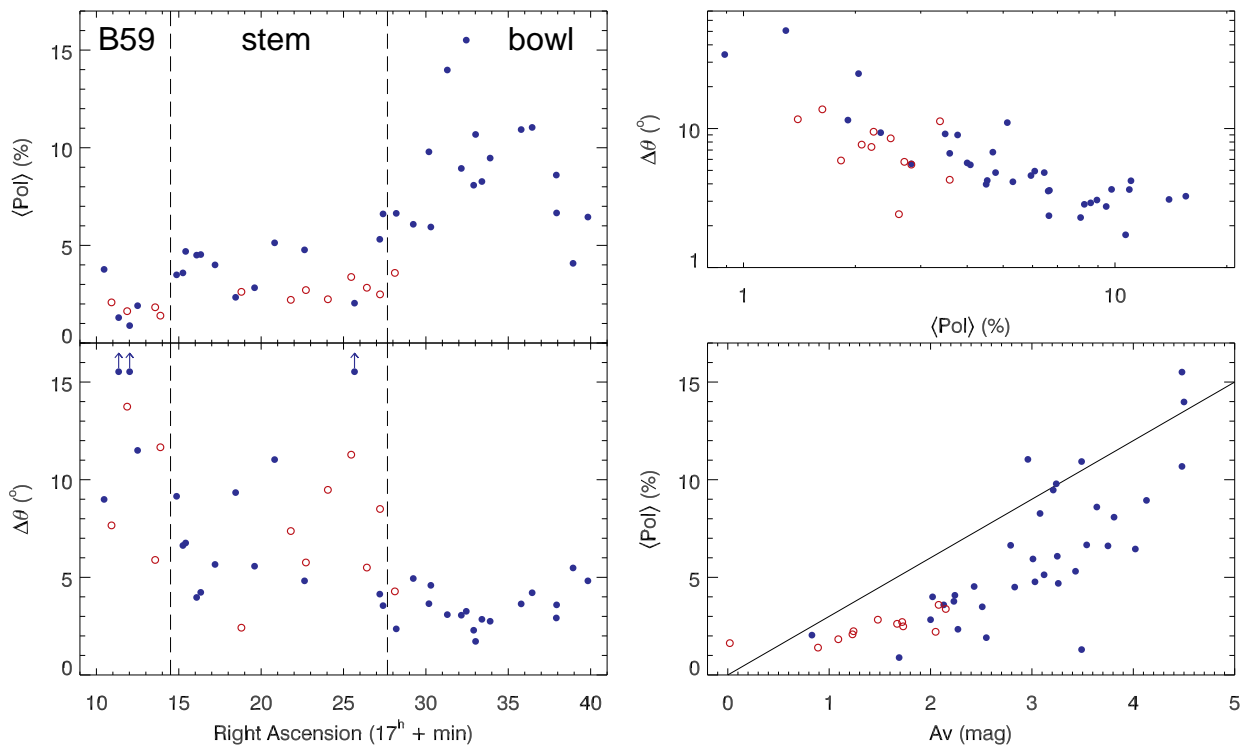}}
\caption[]{{\it left panels:} Distribution of the mean polarization and of 
the polarization angle dispersion, $\Delta\theta$, as a function of the 
right ascension of the observed areas, respectively. The polarization angle 
dispersion is corrected by its mean error (i.e.,
$\Delta\theta^2=\sigma^2_{std} - \langle \sigma_\theta \rangle^2$).
The vertical dashed--lines delimits the transition between regions
with different polarimetric properties. Filled and open dots
represent values for fields with and without associated dense
cores, respectively.  As shown by the {\it botton right panel}, the regions 
traced by the optical polarimetry have extinction of ${\rm A}_V \la 2.2$\,mag 
for fields without cores, while the ones associated with cores show
$0.8 \la {\rm A}_V \la 4.5$\,mag. {\it top right panel:} Correlation 
between dispersion in polarization angle and mean polarization. 
{\it botton right panel:} Mean polarization versus visual absorption 
derived from the 2MASS data for the observed stars with 
${\rm P}/\sigma_P \ge 10$. The solid line represents optimum alignment
efficience ($P(\%) = 3 \times A_V$).}
\label{fig2}
\end{figure*}

\section{Polarization at the Pipe nebula}

To analyze the polarization pattern in the Pipe nebula, we estimated the mean
polarization and position angle for each observed field. To improve the 
precision of the mean values, we selected those objects with 
${\rm P}/\sigma_P \ge 10$ and observed polarization angle $\theta_{obs}$ 
within the interval $(\theta_{av} - 2\sigma_{std}) \le \theta_{obs} \le 
(\theta_{av} + 2\sigma_{std})$ where, $\theta_{av}$ and $\sigma_{std}$ 
are the mean polarization angle and standard deviation of each field 
sample, respectively. We then estimated the mean Stokes parameters for 
each field, from the individual values for each star weighted by the 
estimated observational error. Most fields show a distribution of
polarization position angles that resembles a normal distribution, although 
a more complex distribution is evident in some directions. A detailed 
analysis of these distributions is beyond the scope of the present Letter 
and will be presented in the aforementioned paper. 

Figure~\ref{fig1} shows the mean polarization vectors overlaid on the 2MASS 
infrared extinction map of the Pipe nebula derived by \citet{LAL06}. For most 
fields, the values of the mean polarization and position angle were obtained 
from samples of more than 100 stars. The high signal-to-noise ratio of our 
data set ensures good statistics in our analyses and implies that the degree 
of polarization measured for most fields and, in particular, the significant 
range of mean polarization values derived along the Pipe (from 1 to 15\%)
are truly remarkable. It is also remarkable that the polarization position 
angle does not change significantly along the 17~pc extent of the Pipe nebula 
covered by our observations ($\langle \theta \rangle \simeq 
160\degr$-$10\degr$ for 37 of the 46 fields, where the mean position angles
are given in equatorial coordinates, measured from north to east).  Although 
the physical processes involved in grain alignment is a debated issue 
\citep[see ][for a comprehensive review on this subject]{L03}, it is widely 
believed that starlight polarization is caused by the alignment of elongated 
dust grains by the magnetic field, as suggested by the pioneering work of
\citet{DG51}. Based on this assumption, the polarization map showed in 
Fig.\,\ref{fig1} provides an outline of the magnetic field component parallel 
to the plane of the sky. The almost perpendicular alignment between the 
magnetic field and the main axis of the Pipe's stem is clearly evident. 

It is instructive to analyze the behavior of polarization properties 
along the Pipe nebula: the {\it left panels} of Figure~\ref{fig2} present 
the distribution of the mean polarization and the polarization angle 
dispersion as a function of the right ascension of the observed areas, which 
runs almost parallel to the main axis of the Pipe's stem. Since the 
polarization properties of each field are inhomogeneous, a global analysis 
allows one to distinguish three regions throughout the cloud with rather 
different features between them. These regions, separated by dashed-lines in 
Fig.\,\ref{fig1} and \ref{fig2}, can be identified as: the B59 region, at 
the northwestern end of the cloud; the main filamentary structure (the stem 
of the Pipe); and the irregular--shaped gas at the other extreme end 
(the ``bowl"). We note that fields without cores (open dots in 
Fig.\,\ref{fig2}) show a smaller variation in polarization properties than 
fields with cores (filled dots). 
 
The lowest mean polarizations are observed in the vicinity of B59, the 
only place in the Pipe with evidence of star formation. Seven out of 
eight observed fields in this region show mean polarization degrees of
around 1--2\%. This region has a large polarization angle dispersion. 
Indeed, two fields have a dispersion in polarization angles 
$\Delta\theta \ga 25\degr$ -- these are indicated in the {\it botton right 
panel} of Fig.\,\ref{fig2} by the arrowed dots -- and show the lowest 
mean degree of polarization among the observed fields. We point 
out that the field showing the highest dispersion in polarization angles 
($\Delta\theta \simeq 51\degr$) has a line of sight passing close to the 
densest core of B59, the most opaque region of the Pipe \citep{RZ07}, 
and that our sample has only 12 stars for which ${\rm P}/\sigma_P \ge 10$.

Toward the stem region the mean polarization rises a few percent and
the polarization angle dispersion decreases slightly with respect to B59. 
Most fields containing dense cores show a mean polarization degree 
($\simeq$3--5\%) that is higher than fields without cores ($\simeq$2--3\%)). 
However, this difference is unclear from the position angle dispersion values,
which show a large range of values for both types of field 
($\Delta\theta \simeq$3$\degr$--$12\degr$).

The bowl has a significantly different mean polarization and dispersion 
in position angles: for this region, we measure the highest degree of 
polarization and the lowest dispersion in position angles. Most observed 
fields in the bowl shows a mean polarization higher than about 8\% 
(up to 15\%) and a dispersion in polarization angles of less than $5\degr$. 
This part of the cloud has the most precise alignment between the mean 
polarization vectors of neighboring fields. The high polarization degree 
in the bowl is unusual, since the polarization degree of this type 
of dark interstellar clouds is typically 1 order of magnitude lower than 
we measure and rarely reaches such high values \citep[e.g., ][for optical 
polarimetric data on $\rho$ Oph, Chamaeleon I, and Taurus dark clouds, 
respectively]{VC93, WG94, WG01}. Such a result implies a high efficiency 
of grain alignment for the interstellar dust in those fields, and
that the magnetic field in the bowl is aligned close to the plane of the sky 
(otherwise the efficiency would be even higher).

Figure \ref{fig2} ({\it top right panel}) also indicates the distribution of 
$\Delta\theta$ as a function of the mean polarization: it is a clear 
observational fact for the observed fields that the higher the 
mean polarization, the lower the dispersion in polarization angles. 
The anti-correlation between the dispersion in polarization angles 
and polarization degree has a similar dependence for fields with and
without cores. This anti-correlation could be due just to  projection 
effects: the magnetic field direction changes along the Pipe nebula. However, 
this scenario would imply a polarization efficiency and a magnetic field 
strength (see below) that would be unusually high over the entire nebula. 
The star formation activity in B59 probably precludes this scenario. 

What can the aforementioned polarization properties tell us about the
magnetic field in the Pipe nebula? Our dispersion in polarization angles can 
be used to estimate the magnetic field strength for the observed fields from 
the modified Chandrasekhar-Fermi formula \citep{CF53, OSG01}. The volume 
density and line width of the molecular line emission associated with the dust 
that produces the observed optical polarization and extinction 
can be estimated from the molecular data available in the literature.
Thus, extrapolating the median volume density of cores given by 
\citet{LMRAL} to the  optical polarization zone (which is typically at 
a distance of about  $5\arcmin$ -- 0.2~pc in projection -- from the 
center of the cores), we obtain a volume density of 
$n({\rm H}_2)\simeq3\times10^3$~cm$^{-3}$. We also adopt the line width 
found for C$^{18}$O toward the cores in B59 and the stem, 
0.4~km~s$^{-1}$, and the bowl, 0.5~km~s$^{-1}$ \citep[the values used
here are the ones given by ][]{Mu07}. Assuming these values, we find that the 
magnetic field strength in the B59 region, stem, and bowl, in the plane of the
sky, are about 17, 30, and 65~$\mu$G, respectively (the uncertainty in the 
values are probably less than a factor of 2). Adopting a mean visual 
extinction of 3~mag for the molecular cloud traced by the optical 
polarimetry, we find that the mass--to--flux ratio is about 1.4 (slightly 
super-critical) for B59, in contrast to 0.8 and 0.4 (sub-critical) for the 
stem and the bowl, respectively.

The almost perpendicular alignment between the magnetic field and the main 
axis of the Pipe nebula's stem indicates clearly that this part of the 
cloud contracted in the direction of the field lines. This agrees with 
predictions of the ambipolar-diffusion driven model, for which the first 
evolutionary stage of a typical cloud is dynamical relaxation along field 
lines, almost without lateral contraction, until a quasi-equilibrium state 
is reached \citep[e.g., ][]{FM93, TK07}. Indeed, the magnetic pressure 
($P_{mag} = B^2/8\pi$) of the diffuse part of the cloud (where is most of 
the mass) is the dominant source of pressure in the direction perpendicular 
to the field lines ($12 \times 10^5$ and $2.6 \times 10^5$\,K\,cm$^{-3}$ for 
the bowl and stem, respectively), being higher than the pressure due to the 
weight of the cloud \citep[$P_{cloud}/k = 10^5$\,K\,cm$^{-3}$, 
according to ][]{LMRAL}. This can explain the clear elongated structure 
perpendicular to the magnetic field of the whole nebula.

The derived mean polarization degree and dispersion in polarization angles 
are consistent with a scenario in which the B59 region, the stem, and the bowl 
are  experiencing different stages of their evolution. The weak magnetic 
field derived for the B59's neighboring appears to be the reason for 
it being the only known active star-forming site in the cloud. 
Following the evolutionary sequence, the stem with a mass--to--flux ratio 
close to unity would be the part of the cloud in a transient evolutionary 
state, which is experiencing ambipolar diffusion but has not yet given birth 
to stars.  Finally, the high polarization degree of the bowl combined with 
the low dispersion in the mean polarization vectors implies that the magnetic 
field in this part of the cloud has a major role in regulating the collapse 
of the cloud material compared to the other parts. This would imply that the 
bowl is in a primordial evolutionary state (in the sub-critical regime), not 
yet flattened neither elongated. However, the presence of multiple and 
clearly evident cores implies that fragmentation is already occurring inside
the bowl. A similar case, in a more evolved state, appears to be the Taurus 
molecular cloud complex \citep{NL08}.

\section{Conclusions}

We have described the global polarimetric properties of the Pipe nebula as 
an increasing polarization degree along the filamentary structure from B59 
towards the bowl, while the dispersion in polarization angles decreases 
along this way. Our results appears to indicate that there exist three 
regions in the Pipe nebula of distinct evolutionary stages: since 
the mean orientation angle of the mean polarization vectors is 
perpendicular to the longer axis of the cloud, this implies that the 
cloud collapse is taking place along the magnetic field lines. We
can subdivide the Pipe nebula into the following components:  

\begin{itemize}
\item B59, the only active star-forming site in the cloud. For the observed 
fields, we measure a large dispersion in polarization angle and low 
polarization degree.   
\item The stem, which collapsed by means of ambipolar diffusion but has not 
yet given birth to stars. It appears to represent a transient evolutionary 
state between B59 and the bowl.
\item The bowl, which contains the fields of the highest values of mean 
polarization and the lowest values of dispersion in polarization angle. These 
values imply that the dust grains in the bowl are highly aligned by a 
rather strong magnetic field. For this reason, the bowl may represent the
start of the contraction phase during a very early evolutionary stage. 
\end{itemize} 

\acknowledgements{We thank the staff of the Observat\'orio do Pico dos
Dias (LNA/MCT, Brazil) for their hospitality and invaluable help during
our observing runs. Drs. A. M. Magalh\~aes and A. Pereyra are
acknowledged for providing the polarimetric unit and the software used
for data reductions. 
We made use of NASA's Astrophysics Data System (NASA/ADS) and 
of data products from the Two Micron All Sky Survey.
This research has been partially supported by
CEX APQ-1130-5.01/07 (FAPEMIG, Brazil) and AYA2005--08523--C03
(Ministerio de Ciencia e Innovaci\'on, Spain).}


\begin{thebibliography}{28}
\expandafter\ifx\csname natexlab\endcsname\relax\def\natexlab#1{#1}\fi

\bibitem[{{Alves} \& {Franco}(2007)}]{AF07}
{Alves}, F.~O. \& {Franco}, G.~A.~P. 2007, \aap, 470, 597

\bibitem[{{Alves} {et~al.}(2007){Alves}, {Lombardi}, \& {Lada}}]{ALL07}
{Alves}, J., {Lombardi}, M., \& {Lada}, C.~J. 2007, \aap, 462, L17

\bibitem[{{Brooke} {et~al.}(2007){Brooke}, {Huard}, {Bourke}, {Boogert},
  {Allen}, {Blake}, {Evans}, {Harvey}, {Koerner}, {Mundy}, {Myers}, {Padgett},
  {Sargent}, {Stapelfeldt}, {van Dishoeck}, {Chapman}, {Cieza}, {Dunham},
  {Lai}, {Porras}, {Spiesman}, {Teuben}, {Young}, {Wahhaj}, \& {Lee}}]{BHB07}
{Brooke}, T.~Y., {Huard}, T.~L., {Bourke}, T.~L., {et~al.} 2007, \apj, 655, 364

\bibitem[{{Chandrasekhar} \& {Fermi}(1953)}]{CF53}
{Chandrasekhar}, S. \& {Fermi}, E. 1953, \apj, 118, 113

\bibitem[{{Davis} \& {Greenstein}(1951)}]{DG51}
{Davis}, L.~J. \& {Greenstein}, J.~L. 1951, \apj, 114, 206

\bibitem[{{Elmegreen} \& {Scalo}(2004)}]{ES04}
{Elmegreen}, B.~G. \& {Scalo}, J. 2004, \araa, 42, 211

\bibitem[{{Fiedler} \& {Mouschovias}(1993)}]{FM93}
{Fiedler}, R.~A. \& {Mouschovias}, T.~C. 1993, \apj, 415, 680

\bibitem[{{Franco} {et~al.}(2008){Franco}, {Alves}, \& {Girart}}]{FAG08}
{Franco}, G.~A.~P., {Alves}, F.~O., \& {Girart}, J.~M. 2008, in preparation

\bibitem[{{Goodman} {et~al.}(1990){Goodman}, {Bastien}, {Menard}, \&
  {Myers}}]{GBM90}
{Goodman}, A.~A., {Bastien}, P., {Menard}, F., \& {Myers}, P.~C. 1990, \apj,
  359, 363

\bibitem[{{Lada} {et~al.}(2008){Lada}, {Muench}, {Rathborne}, {Alves}, \&
  {Lombardi}}]{LMRAL}
{Lada}, C.~J., {Muench}, A.~A., {Rathborne}, J., {Alves}, J.~F., \& {Lombardi},
  M. 2008, \apj, 672, 410

\bibitem[{{Lazarian}(2003)}]{L03}
{Lazarian}, A. 2003, Journal of Quantitative Spectroscopy and Radiative
  Transfer, 79, 881

\bibitem[{{Lizano} \& {Shu}(1989)}]{LS89}
{Lizano}, S. \& {Shu}, F.~H. 1989, \apj, 342, 834

\bibitem[{{Lombardi} {et~al.}(2006){Lombardi}, {Alves}, \& {Lada}}]{LAL06}
{Lombardi}, M., {Alves}, J., \& {Lada}, C.~J. 2006, \aap, 454, 781

\bibitem[{{Mac Low} \& {Klessen}(2004)}]{MK04}
{Mac Low}, M.-M. \& {Klessen}, R.~S. 2004, Reviews of Modern Physics, 76, 125

\bibitem[{{Magalh\~aes} {et~al.}(1996){Magalh\~aes}, {Rodrigues}, {Margoniner},
  {Pereyra}, \& {Heathcote}}]{AM96}
{Magalh\~aes}, A.~M., {Rodrigues}, C.~V., {Margoniner}, V.~E., {Pereyra}, A.,
  \& {Heathcote}, S. 1996, in ASP Conf. Ser. 97: Polarimetry of the
  Interstellar Medium, 118

\bibitem[{{Mestel} \& {Spitzer}(1956)}]{MS56}
{Mestel}, L. \& {Spitzer}, Jr., L. 1956, \mnras, 116, 503

\bibitem[{{Mouschovias} \& {Paleologou}(1981)}]{MP81}
{Mouschovias}, T.~C. \& {Paleologou}, E.~V. 1981, \apj, 246, 48

\bibitem[{{Mouschovias} {et~al.}(2006){Mouschovias}, {Tassis}, \&
  {Kunz}}]{MTK06}
{Mouschovias}, T.~C., {Tassis}, K., \& {Kunz}, M.~W. 2006, \apj, 646, 1043

\bibitem[{{Muench} {et~al.}(2007){Muench}, {Lada}, {Rathborne}, {Alves}, \&
  {Lombardi}}]{Mu07}
{Muench}, A.~A., {Lada}, C.~J., {Rathborne}, J.~M., {Alves}, J.~F., \&
  {Lombardi}, M. 2007, \apj, 671, 1820

\bibitem[{{Nakamura} \& {Li}(2008)}]{NL08}
{Nakamura}, F. \& {Li}, Z.-Y. 2008, arXiv:0804.4201v1 [astro-ph]

\bibitem[{{Nakano}(1979)}]{Na79}
{Nakano}, T. 1979, \pasj, 31, 697

\bibitem[{{Ostriker} {et~al.}(2001){Ostriker}, {Stone}, \& {Gammie}}]{OSG01}
{Ostriker}, E.~C., {Stone}, J.~M., \& {Gammie}, C.~F. 2001, \apj, 546, 980

\bibitem[{{Rom{\'a}n-Z{\'u}{\~n}iga} {et~al.}(2007){Rom{\'a}n-Z{\'u}{\~n}iga},
  {Lada}, {Muench}, \& {Alves}}]{RZ07}
{Rom{\'a}n-Z{\'u}{\~n}iga}, C.~G., {Lada}, C.~J., {Muench}, A., \& {Alves},
  J.~F. 2007, \apj, 664, 357

\bibitem[{{Tassis} \& {Mouschovias}(2004)}]{TM04}
{Tassis}, K. \& {Mouschovias}, T.~C. 2004, \apj, 616, 238

\bibitem[{{Tassis} \& {Mouschovias}(2007)}]{TK07}
{Tassis}, K. \& {Mouschovias}, T.~C. 2007, \apj, 660, 388

\bibitem[{{Vrba} {et~al.}(1993){Vrba}, {Coyne}, \& {Tapia}}]{VC93}
{Vrba}, F.~J., {Coyne}, G.~V., \& {Tapia}, S. 1993, \aj, 105, 1010

\bibitem[{{Whittet} {et~al.}(1994){Whittet}, {Gerakines}, {Carkner}, {Hough},
  {Martin}, {Prusti}, \& {Kilkenny}}]{WG94}
{Whittet}, D.~C.~B., {Gerakines}, P.~A., {Carkner}, A.~L., {et~al.} 1994,
  \mnras, 268, 1

\bibitem[{{Whittet} {et~al.}(2001){Whittet}, {Gerakines}, {Hough}, \&
  {Shenoy}}]{WG01}
{Whittet}, D.~C.~B., {Gerakines}, P.~A., {Hough}, J.~H., \& {Shenoy}, S.~S.
  2001, \apj, 547, 872

\end{thebibliography}
\end{document}